\documentclass[sigconf]{acmart}

\usepackage{graphicx} 
\usepackage{algorithmic}
\usepackage{textcomp}
\usepackage{soul}
\usepackage{color}
\usepackage{xcolor}
\usepackage{xspace}
\usepackage{geometry}
\usepackage{array}
\usepackage{threeparttable}
\usepackage{enumitem}
\usepackage{makecell}
\usepackage{multirow}
\usepackage{colortbl}
\usepackage{paralist}

\newcommand{\botium}{\textsc{Botium}\xspace}
\newcolumntype{C}[1]{>{\centering\arraybackslash}m{#1}}

\copyrightyear{2024}
\acmYear{2024}
\setcopyright{rightsretained}
\acmConference[ICSE-Companion '24]{2024 IEEE/ACM 46th International
Conference on Software Engineering: Companion Proceedings}{April 14--20,
2024}{Lisbon, Portugal}
\acmBooktitle{2024 IEEE/ACM 46th International Conference on Software
Engineering: Companion Proceedings (ICSE-Companion '24), April 14--20, 2024,
Lisbon, Portugal}\acmDOI{10.1145/3639478.3640032}
\acmISBN{979-8-4007-0502-1/24/04}

\begin{document}

\title{\approach: A Mutation Testing Approach for Chatbots}

\author{Michael Ferdinando Urrico}
\email{m.urrico@campus.unimib.it}
\affiliation{%
  \institution{University of Milano-Bicocca}
  \city{Milan}
  \country{Italy}
}
\author{Diego Clerissi}
\email{diego.clerissi@unimib.it}
\affiliation{%
  \institution{University of Milano-Bicocca}
  \city{Milan}
  \country{Italy}
}
\author{Leonardo Mariani}
\email{leonardo.mariani@unimib.it}
\affiliation{%
  \institution{University of Milano-Bicocca}
  \city{Milan}
  \country{Italy}
}

\renewcommand{\shortauthors}{Urrico et al.}

\newcommand{\approach}{\textsc{MutaBot}\xspace}

\newcommand{\numChatbots}%
{three\xspace}

\newcommand{\numOperators}{24\xspace}

\begin{abstract}
Mutation testing is a technique aimed at assessing the effectiveness of test suites by seeding artificial faults into programs. Although available for many platforms and languages, no mutation testing tool is currently available for conversational chatbots, which represent an increasingly popular solution to design systems that can interact with users through a natural language interface. Note that since conversations must be explicitly engineered by the developers of conversational chatbots, these systems are exposed to specific types of faults not supported by existing mutation testing tools.  

In this paper, we present \approach, a mutation testing tool for conversational chatbots. \approach addresses mutations at multiple levels, including conversational flows, intents, and contexts. We designed the tool to potentially target multiple platforms, while we implemented initial support for Google Dialogflow chatbots. We assessed the tool with \numChatbots Dialogflow chatbots and test cases generated with \botium, revealing weaknesses in the test suites.
\end{abstract}

\ccsdesc[500]{Software and its engineering~Software testing and debugging}
\ccsdesc[100]{Human-centered computing~Natural language interfaces}

\keywords{Chatbot Testing, Mutation Testing, Botium, Dialogflow}


\maketitle

\section{Introduction}\label{sec:introduction}The widespread adoption of conversational chatbots (e.g., 88\% of people had at least a conversation with a chatbot in 2022~\cite{chatbots-stats}), also known as virtual assistants or conversational agents, in various domains (e.g., e-commerce, booking, tech support, healthcare, and more)~\cite{adamopoulou2020chatbots,shawar2007chatbots,grudin2019chatbots} raised the need of dedicated testing methodologies and tools~\cite{botium,bespoken,qbox,ruane2018botest,guichard2019assessing,bravo2020testing,bozic2019chatbot,bovzic2022ontology} to assess their effectiveness.  

Mutation testing is a popular validation strategy that can be used to assess the effectiveness of test suites and test case generation techniques~\cite{jia2010analysis}. It is a fault-based testing technique that consists of producing many faulty versions of a software under test, called \textit{mutants}, by systematically introducing artificial faults in the software. The faults are introduced through \textit{mutation operators}, designed to change the software according to patterns that represent possible faults. For instance, an operator may replace the name of a variable in an expression with another one to represent the case of a wrongly referred variable.  
The effectiveness of test suites is assessed by measuring the rate of mutants that they can reveal, that is, the percentage of mutants that fail (i.e., \textit{killed}, according to the mutation testing terminology~\cite{jia2010analysis}) when executed with the test suite. 

In order to apply mutation testing in a given context, it is necessary to define a set of operators that can modify the relevant language elements and entities. Conversational chatbots, differently from other software, implement a number of components with a specific structure, even spanning multiple diverse languages, to encode the supported conversations. For instance, chatbots implemented with Google Dialogflow~\cite{dialogflow} are composed of several JSON files that specify both the sentences that a chatbot can use and the possible flows of conversation that the chatbot can follow, text files with alternative sentences, and webhooks to trigger external functions. This case is the same for all the chatbot platforms (e.g., Rasa~\cite{rasa}  and Amazon Lex~\cite{amazon-lex}), where conversations and interactions are implemented with specific language elements. 

Currently, there is \emph{no mutation testing tool that can properly target the elements that implement the conversations}, taking the semantics of the mutated element into consideration. On the other hand, conversations need to be carefully validated since they are the main and only interface used by chatbots to interact with the outside world.


 \begin{figure*}[t!]
 \centering
 \includegraphics[trim=0.0cm 0.0cm 0.0cm 0.0cm, clip=true, width=17.5cm]{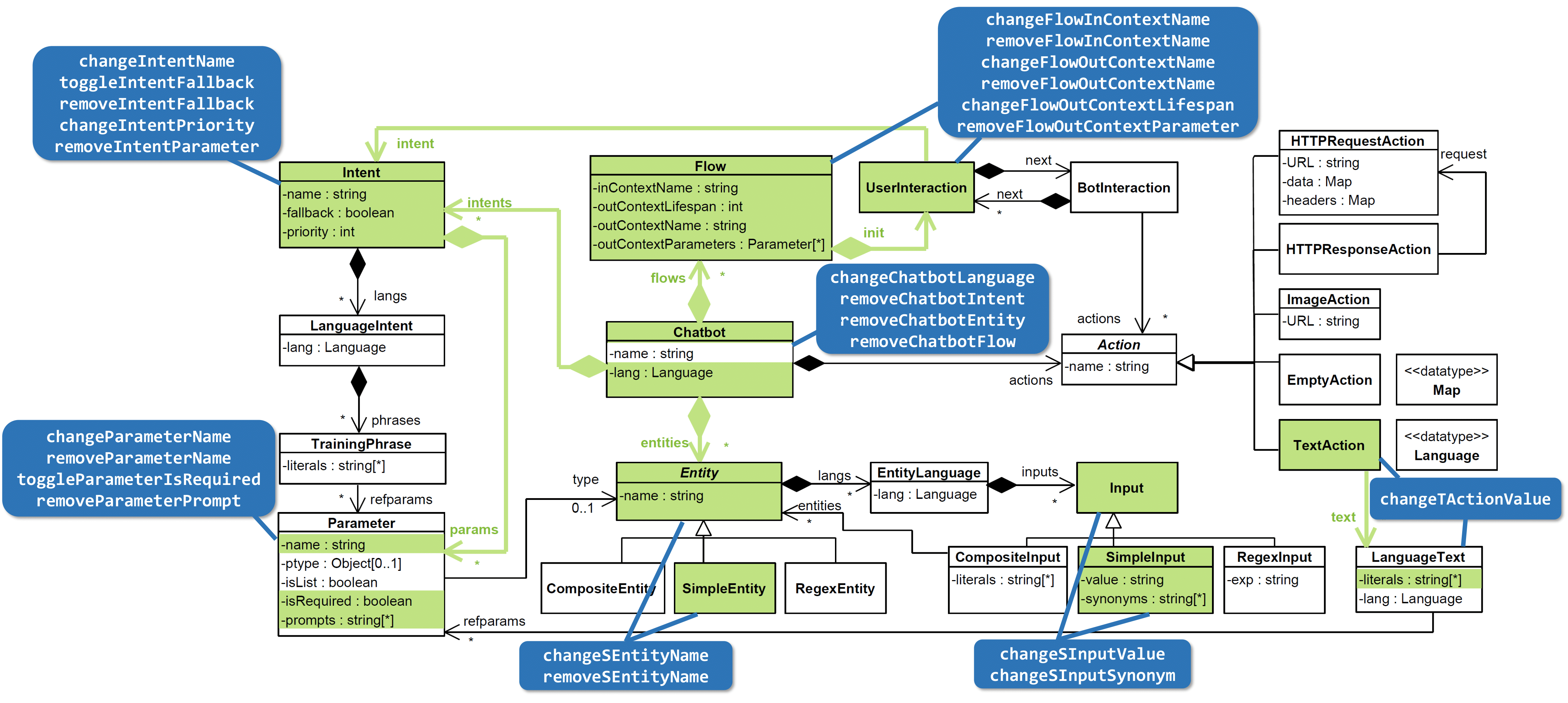}
 \caption{Chatbot structure meta-model adapted from Ca{\~n}izares \textit{et al.}~\cite{canizares2022automating}.}
 \label{fig:meta-model}
 \end{figure*}

In this paper, we introduce \approach, to be best of our knowledge the first mutation testing tool for chatbots. \approach is characterized by 
a general-purpose technology-agnostic  design, inspired by the meta-model proposed by Ca{\~n}izares \textit{et al.}~\cite{canizares2022automating}, that can be exploited to virtually address any chatbot platforms. When a specific platform has to be addressed, the technology-dependent interfaces, which provide the operations to locate the items to be mutated in a given platform, have to be implemented. Our first prototype version of the tool implements \numOperators mutation operators specifically designed for chatbots and provides support to Google Dialogflow~\cite{dialogflow}, which is among the most popular chatbot platforms.

We preliminary validated our tool with \numChatbots Dialogflow chatbots, applying all the supported mutations, and observed how the test suites generated with \botium~\cite{botium}, a state-of-the-art automated test generation framework for chatbots, perform in detecting mutants. Results show that the test suites were capable of killing only between 28\% and 37\% of the mutants, indicating how test case generation tools for chatbots need to be improved to better cover the space of the possible conversations.

The paper is organized as follows. Section~\ref{sec:tool} describes \approach, its architecture and the supported mutation operators. Section~\ref{sec:evaluation} presents the procedure we followed to conduct our empirical evaluation and discusses the results. Section~\ref{sec:related} discusses the related work. Finally, Section~\ref{sec:conclusion} provides final remarks.

\section{\approach Approach}\label{sec:tool}
The mutation operators implemented in \approach are inspired by the chatbot meta-model originally proposed by Ca{\~n}izares \textit{et al.}~\cite{canizares2022automating}, which specifies general platform-agnostic chatbot concepts and features, derived by analyzing 15 chatbot development platforms~\cite{perez2021creating}. 

Figure~\ref{fig:meta-model} shows the meta-model. A \textit{Chatbot} is represented as a class composed of \textit{Intents} (i.e., the possible goals of a user), \textit{Entities} (i.e., the data types that can be used in a conversation), \textit{Actions} (i.e., the possible responses of a chatbot), and conversational \textit{Flows} (i.e., the possible user-bot interaction flows). 
As chatbots are intended to be multi-language, \textit{Intents} declare the \textit{TrainingPhrases} used to train the chatbot for each supported language. \textit{TrainingPhrases} can use \textit{Parameters}, typed according to the supported entities. \textit{Entities} can be \textit{Simple} (i.e., a literal with synonyms), \textit{Complex} (i.e., a composition of other entities and literals), or \textit{Regex} (i.e., values derived from regular expressions).
In response to a user request, a chatbot performs an \textit{Action}, which can be of type \textit{Text} (i.e., a plain text response), \textit{Image} (i.e., a graphical object to interact with, shown to the user), a \textit{HTTPRequest} and \textit{HTTPResponse} (i.e., an action invoking an external service to respond to the user), or \textit{Empty} (i.e., no action is performed). Finally, a conversational \textit{Flow} determines the sequence of interactions between the user and the chatbot, alternatively activating \textit{Intents} and \textit{Actions}. We added the notions of \textit{name},  \textit{lifespan}, and \textit{parameters} of the context data (i.e., the memory of the chatbot) associated with each conversational flow to the meta-model, since these are also important concepts that determine how conversations work in multiple platforms. 

We annotated Figure~\ref{fig:meta-model} to indicate the \numOperators operators currently supported by \approach. We list the name of the operator in a callout text, and we highlight the area of the meta-model affected by the operator. We defined operators targeting all the main entities and features that play a relevant role in a conversation. Unlike traditional data mutation techniques, \approach implements operators carefully designed to manage the complexity of the target platform, to identify the software element exactly corresponding to the conceptual element presents in the meta-model. 

The operators defined in \approach drastically differ from traditional mutation operators affecting statements, such as array index manipulations or  decision mutation operators. Instead, the implemented operators address the peculiarities of conversational agents. In fact, each operator defined on the meta-model may implement changes of different nature depending on the specific chatbot technology. For instance, changing the flow of a conversation requires changing the values stored in the \textit{JSON} contexts objects for DialogFlow, while they can be directly altered by changing \textit{YAML} configuration files in Rasa.    



The implemented operators represent faults that may originate from the imprecise design of the conversations, such as, mistakenly swapping the name of one parameter with another one in a response, forgetting to propagate the context from one intent to another, or not setting a mandatory parameter as required.
We briefly present below the list of supported operators, and the tool.
For a more detailed view of the operators and the tool, please refer to the \approach project repository\footnote{https://gitlab.com/Michael-Urrico/defectinjector-java}.

\subsection{Mutant Operators} \label{sec:mutants}

We list below the mutant operators grouped by the entity that is affected by the operators, according to Figure~\ref{fig:meta-model}.   

\noindent \textbf{Chatbot}:
\begin{inparaenum}
    \item \textit{changeChatbotLanguage}: replace a chatbot supported language with another existing\footnote{To generate more realistic mutations, we defined the semantics of replace operators like \textit{replace * with another existing *} (e.g., \textit{changeChatbotLanguage}) as replacing the value of the target property with another existing value of the same property, rather than using a randomly generated value (e.g., replacing the name of an intent with another existing intent name). We plan to add the generation of random values in the future.} language, 
    \item \textit{removeChatbotIntent}: remove a chatbot intent, 
    \item \textit{removeChatbotEntity}: remove a chatbot entity,
    \item \textit{removeChatbotFlow}: remove a chatbot flow. 
    
\end{inparaenum}


\noindent \textbf{Flow}:
\begin{inparaenum}
    \item \textit{changeFlowInContextName}: replace a flow input context name with another existing name,
    \item \textit{removeFlowInContextName}: remove a flow input context name,
     \item \textit{changeFlowOutContextName}: replace a flow output context name with another existing name,
    \item \textit{removeFlowOutContextName}: remove a flow output context name,
    \item \textit{changeFlowOutContextLifespan}: replace a flow output context lifespan with a random value between 1 and 3, different from the original one,
    \item \textit{removeFlowOutContextParameter}: remove a flow output context parameter. 
    
\end{inparaenum}


\noindent \textbf{Intents}: 
\begin{inparaenum}
    \item \textit{changeIntentName}: replace an intent name with another existing name, 
    \item \textit{toggleIntentFallback}: switch an intent "fallback" flag to true/false,
    \item \textit{removeIntentFallback}: remove an intent "fallback" flag,
    \item \textit{changeIntentPriority}: replace an intent priority with a random value between 0 and 1,000,000 (max priority value) different from the original one,
    \item \textit{removeIntentParameter}: remove an intent parameter.  
    
\end{inparaenum}


\noindent \textbf{Parameters}: 
\begin{inparaenum}
    \item \textit{changeParameterName}: replace a parameter name with another existing name, 
    \item \textit{removeParameterName}: remove a parameter name,
    \item \textit{toggleParameterIsRequired}: switch a parameter "isRequired" flag to true/false,
    \item \textit{removeParameterPrompt}: remove a parameter prompt.
\end{inparaenum}


\noindent \textbf{Inputs}: 
\begin{inparaenum}
    \item \textit{changeSEntityName}: replace an entity name with another existing name,
    \item \textit{removeSEntityName}: remove an entity name,
    \item \textit{changeSInputValue}: replace an input value with a random string,
    \item \textit{changeSInputSynonym}:
    replace an input synonym with a random string,
    \item \textit{changeTActionValue}: replace a textual action value with a random string.
\end{inparaenum}

\begin{figure}[t!]
\centering
\includegraphics[trim=1.0cm 16.5cm 1.0cm 1.0cm, clip=true, width=8.4cm]{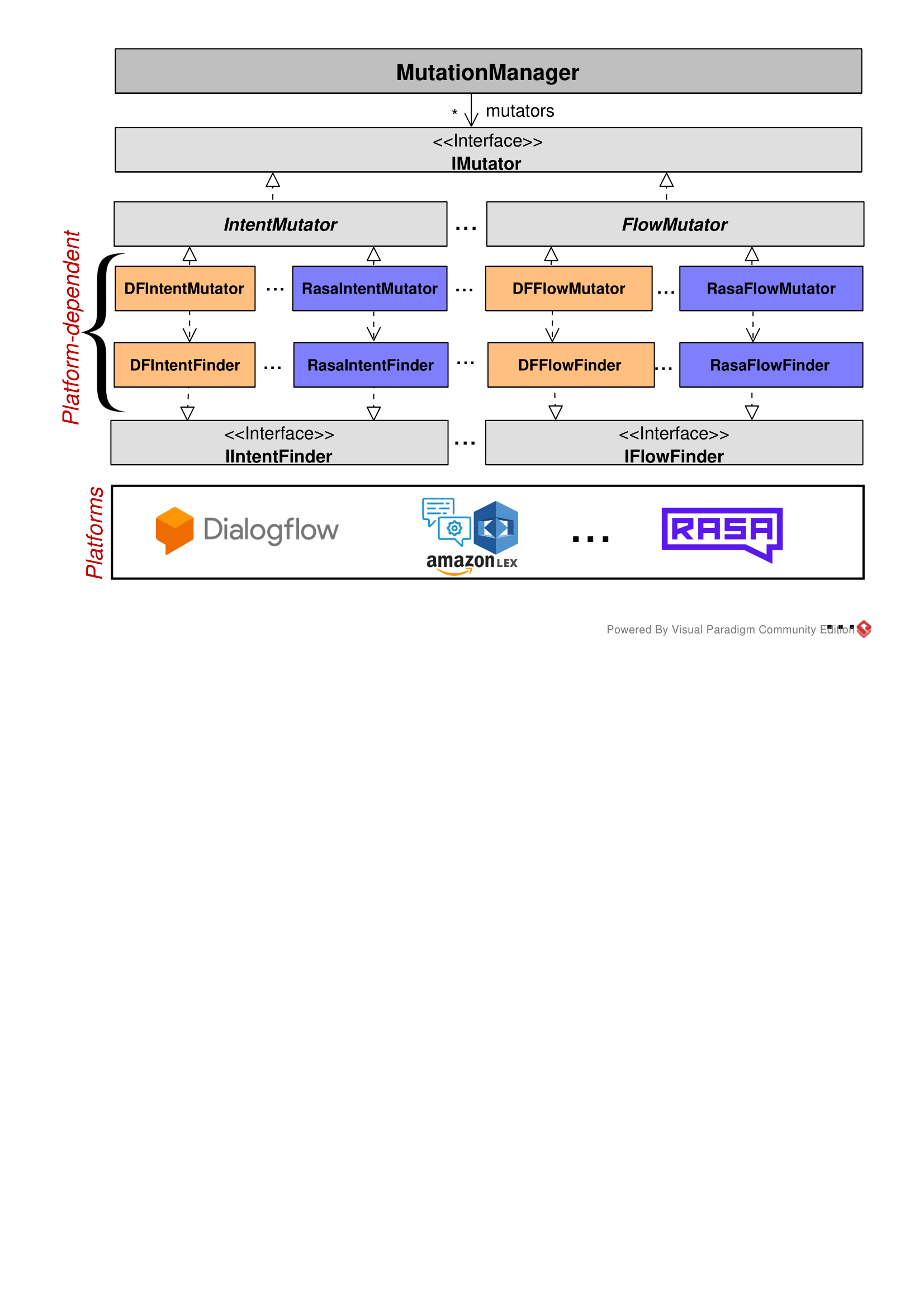}
\caption{\approach architecture.}
\label{fig:architecture}
\end{figure}

\subsection{\approach Tool}

The architecture of the tool is shown in Figure~\ref{fig:architecture}. Its core component is a \texttt{MutationManager} that is responsible of running the individual mutators available in the tool. Each mutator is implemented as an abstract operator, responsible of actuating the change, and an abstract finder, responsible of finding the code element that must be changed by the operator. These abstract components are refined for the target platform. Our prototype supports Dialogflow chatbots. 
In order to find the elements to mutate, \approach knows how to search within the files that compose a chatbot and process them.


The tool takes in input multiple XML configuration files that specify the target folder that hosts the code of the chatbot, the type of desired output (a full copy of the mutated chatbot, or only a copy of the modified files), and the set of operators that must be applied. The output consists of a set of mutated chatbots, either including the full executable code of the bot or only the modified files, and a report file with the list of seeded mutations.



\section{Empirical Evaluation}\label{sec:evaluation}
\begin{table}[t!]
\caption{Subject Chatbots.}
\label{tab:chatbots}
\centering
\footnotesize
\begin{threeparttable}
\begin{tabular}{|C{2.68cm}|C{1.85cm}|C{1.6cm}|C{0.74cm}|}
    \hline
    \textbf{Name} & \textbf{Domain} & \textbf{\# Int./Ent./Act.} & \textbf{\# Tests} \\
    \hline
     AppointmentScheduler~\cite{perez2021creating}\tnote{a} & Booking & 3/3/19 & 21(0)\\ 
     Device~\cite{perez2021creating}\tnote{b} & Home Automation & 11/2/11 & 95(28)\\
    Nutrition~\cite{perez2020model,bravo2020testing}\tnote{c} & Food \& Health  & 4/7/18 & 46(1)\\
    \hline
\end{tabular}
\begin{tablenotes}\footnotesize
    \item[a] https://github.com/priyankavergadia/AppointmentScheduler-GoogleCalendar
    \item[b] https://dialogflow.cloud.google.com/ (prebuilt)
    \item[c] https://github.com/viber/apiai-nutrition-sample
  
\end{tablenotes}
\end{threeparttable}
\end{table}

\noindent We preliminarily evaluated \approach by generating mutants for Dialogflow chatbots, and assessing the effectiveness of the test cases generated with the \botium test case generation tool~\cite{botium}, which is a white-box testing technique developed by Botium GmbH in 2018, widely used in both industrial and academic contexts~\cite{bravo2020testing,li2022review,perez2021choosing}. 
Note that here a test is a conversation, with specific expectations on the responses of the chatbot. The conversations automatically generated by Botium cover the user-bot scenarios that could be derived from the implementation of the chatbot under test. In the case of Dialogflow, conversations are derived from the \textit{JSON} files that encode the possible requests and responses.  

As subjects, we selected \numChatbots third-party Dialogflow chatbots of different domains and features, that have been used in previous studies~\cite{perez2021creating,bravo2020testing,perez2020model}. Table~\ref{tab:chatbots} summarizes their characteristics (i.e., name, domain, and number of intents/entities/actions), and the sizes of the test suites generated by \botium.

For each chatbot, we first ran \botium in \textit{generation} mode in order to generate our baseline regression test suite. We then used \botium in \textit{execution} mode, repeating five times each test suite, in order to detect and discard any spuriously failing test or \textit{flaky test case}\footnote{\textit{"A flaky test is a test that both passes and fails periodically without any code changes.
"}~\cite{zheng2021research}}, that \botium may have generated and that could invalidate our results. We report, between parentheses in Table~\ref{tab:chatbots}, the number of discarded tests.

\begin{table*}[ht!]
\caption{Mutants killed over total by Botium test suites.}
\centering
\resizebox{1.8\columnwidth}{!}{
\begin{tabular}{|c|c|c|c|c|c|>{\columncolor[gray]{0.9}}c|}
\hline
\multirow{2}{*}{\textbf{Chatbots}} & \multicolumn{6}{c|}{\textbf{\# Killed/\# Equivalent/\# Generated (\% Killed)}}\\
& \textbf{Chatbot} & \textbf{Flows} & \textbf{Intents} & \textbf{Parameters} & \textbf{Inputs} & \textbf{Total} \\
\hline
AppointmentScheduler & 4/0/5 (80\%) & - & 4/5/12 (57\%) & 0/0/3 (0\%) & 0/0/7 (0\%) & 8/5/27 (36\%)\\
\hline
Device & 12/0/19 (63\%) & 3/2/12 (30\%) & 11/17/44 (41\%) & 0/0/8 (0\%) & 0/0/6 (0\%) & 26/19/89 (37\%) \\
\hline
Nutrition & 5/2/12 (50\%) & - & 3/7/16 (33\%) & 0/0/6 (0\%) & 1/16/23 (14\%) & 9/25/57 (28\%)\\
\hline
\end{tabular}
}
\label{tab:results}
\end{table*}

We applied \approach to each chatbot, generating mutants with all the operators listed in Section~\ref{sec:mutants}, applying each mutation to every entity instance present in the target chatbot. For instance, the \textit{removeChatbotIntent} operator was applied to every intent present in a tested chatbot, generating mutants each one with a different intent removed. Some mutations were not applicable to all chatbots because of missing properties to mutate (e.g., removing a context parameter from a chatbot with no context).
We finally independently deployed each mutant on Dialogflow and tested it with the \botium test suite. Table~\ref{tab:results} reports the numbers and percentages of \textit{non-equivalent}\footnote{An equivalent mutant is a mutant that cannot be killed by any test case~\cite{jia2010analysis}. We manually inspected the generated mutants to establish their equivalence to the original program, comparing the differences between the original and mutated properties and detecting those that had no impact on the  behavior of the chatbot (e.g., changing the priority of an intent when there are no other intents to compete with).} mutants killed by the test suites, grouped per type of mutated element.

Results show that the test suites were capable of killing only between 28\% and 37\% mutants in total.

Mutations affecting the chatbot structure, like intent or entity removal, as well as those affecting intents properties, are easier to reveal (50\%-80\% for \textbf{Chatbot} category, 33\%-57\% for \textbf{Intents}), with the exception of those targeting the fallback mechanism, which is activated when the bot cannot understand the message of the user. 
The application of \textit{toggleIntentFallback} operator to an intent may introduce tricky behaviors in chatbots, which are sometime detected, but are not always covered by the generated test cases, as they are mainly covering the positive scenarios.
On the other hand, changing the priority of intents mainly produces equivalent mutants, as intents are rarely in competition in terms of which one is supposed to be activated first when a certain user request is formulated.

Mutations targeting the \textbf{Flows} category are  difficult to be revealed (30\% in the Device chatbot), since they require the construction of lengthy and articulated conversations that \botium often fails to create. This group of mutations requires further investigation on more complex chatbots, as the conversational flows cannot be mutated if intents are self-contained.
Finally, the mutations targeting parameters and inputs remain mostly undetected (0\% for the \textbf{Parameters} category, and 14\% at best for \textbf{Inputs}).

These initial results obtained with \approach show how there is still a significant gap to be addressed in terms of the capability of test generation tools to exercise chatbot conversations thoroughly.

\section{Related Work}\label{sec:related}Conversational chatbots are increasingly  exploited in various contexts (e.g., Web, Mobile) to support users 24/7 in a plethora of activities (e.g., booking, home banking, etc.)~\cite{adamopoulou2020chatbots,adamopoulou2020overview}. 
Despite their increasing level of adoption, the dedicated quality assurance solutions are still  limited~\cite{li2022review,bravo2020testing}. Indeed, they are often restricted to individual modules (e.g., the automated speech recognition systems~\cite{iwama2019automated,qin2019imperceptible,asyrofi2020crossasr}),  or they heavily rely on human knowledge and fixtures~\cite{vasconcelos2017bottester,bozic2019chatbot,bovzic2022ontology}, or are limited to  performance evaluation metrics~\cite{deriu2021survey,qbox,chatbottest}.  

Support to test automation mainly consists of frameworks for the automation of test execution. They include solutions to test Alexa skills~\cite{alexa-simulator}, Rasa chatbots~\cite{rasa}, and to generically address end-to-end conversational agents~\cite{playwright,bespoken}. 
The commercial tool \botium~\cite{botium} represents an initial effort in the design of automated functional testing tools for chatbots. Input mutation is further addressed in  Charm~\cite{bravo2020testing}, and
by Guichard \textit{et al.}~\cite{guichard2019assessing}, by building  paraphrases of user queries generated by another tool~\cite{ruane2018botest}. Bozic \textit{et al.} face the oracle problem of chatbot testing via metamorphic relations, involving input transformations, such as synonyms replacements and words removal~\cite{bovzic2022ontology}. 


The lack of mutation testing solutions for conversational chatbots limits the number of faults against which techniques can be assessed, generating difficulties with large scale experimentation. To address this gap, this paper proposes \approach, a mutation testing tool specifically designed to target the conversations implemented by chatbots.

\section{Conclusion}\label{sec:conclusion}
Mutation testing is an important solution to extensively validate test suites and testing techniques, as it ensures that the software under test is thoroughly exercised. So far, no mutation testing tool that can target the conversations implemented by chatbots have been defined, hindering the opportunity to systematically validate test suites against a large number of faults.

In this paper, we presented \approach, a mutation testing tool for chatbots. Our tool targets Google Dialogflow chatbots and implements operators to alter the main elements that may impact a conversation, including intents, conversational flows, entities, and contexts. We reported an initial validation of our tool that shows how the state-of-the-art \botium test generation technique may fail at revealing several mutations, calling for solutions that can more thoroughly validate conversations.

As future work, we aim at exploiting our platform-agnostic architecture to support additional chatbot platforms, such as Amazon Lex and Rasa, and to cover additional elements of the meta-model. Further, we plan to add more variants of the existing operators (e.g., adding operators that mutate properties with random values), and to finally exploit \approach to perform large scale fault-based assessment of the existing test case generation tools for chatbots.




\begin{acks}
This work has been partially funded by the Engineered MachinE Learning-intensive IoT systems (EMELIOT) national research project, which has been funded by the MUR under the PRIN 2020 program (Contract nr. 2020W3A5FY).
\end{acks}

\balance


\end{document}